\documentclass[epj]{svjour}

\usepackage{longtable}
\usepackage{graphicx}

\usepackage{epsfig}
\usepackage{graphics}
\usepackage{latexsym}
\usepackage{amsmath}
\usepackage{amssymb}

\usepackage{amsfonts}
\usepackage{amsmath}
\usepackage{amssymb}
\usepackage{epsfig}
\usepackage{graphicx}

\newcommand{\FP}{\mathcal J}
\newcommand{\bs}{\boldsymbol}
\newcommand{\ar}{\arrowvert}

\newcommand{\be}{\begin{equation}}
\newcommand{\ee}{\end{equation}}
\newcommand{\ba}{\begin{eqnarray}}
\newcommand{\ea}{\end{eqnarray}}

\begin{document}

\title{Meson and tetra-quark mixing}
\author{Ping Wang\inst{1,2}, Stephen R. Cotanch\inst{1} and Ignacio J. General\inst{3}}
\institute{Department of Physics, North Carolina State University,
Raleigh, NC 27695-8202, USA \and Jefferson Laboratory, 12000 Jefferson Ave., Newport News, VA 23606, USA
\and Bayer School of Natural and Environmental Sciences,
Duquesne University, Pittsburgh, PA 15282, USA}
\date{Received: date / Revised version: date}

\abstract{ The mixing between $q \bar{q}$ meson and $q \bar{q} q \bar{q}$ tetra-quark states is examined within an effective
QCD Coulomb gauge Hamiltonian model. Mixing matrix elements of the Hamiltonian are computed and then diagonalized yielding an improved prediction for the low-lying $J^{PC} = 0^{\pm+}, 1^{--}$ isoscalar spectra. Mixing effects were found significant for the scalar hadrons but not for the $1^{--}$ states,   which is consistent with the ideal mixing
of vector mesons.  A  perturbative  assessment of the exact QCD kernel is also reported.}
\PACS{
      {12.39.Mk}{Glueball and exotic multi-quark/gluon states}   \and
      {12.39.Pn}{Potential models}   \and
      {12.39.Ki}{Relativistic quark model}   \and
      {12.40.Yx}{Hadron mass models and calculations}
     } 
%
\maketitle

\section{Introduction}

Hadronic structure remains an interesting but challenging problem.  This is because quantum chromodynamics [QCD] permits
a variety of hadron formations such as  $q \bar{q}$ mesons, $q \bar{q} q \bar{q}$
tetra-quarks, $q \bar{q}g$ hybrids and $gg$ glueballs.  These states will, in general, also mix
via   quark pair annihilation/formation which further complicates this issue.
Unfortunately experimental information \cite{PDG}
is predominantly limited to  masses, widths and spectroscopic quantum numbers with little structure insight.
Accordingly, theoretical input is needed and this paper reports a consistent, model study
of the dynamic mixing between $q \bar{q}$ meson and $q \bar{q} q \bar{q}$ tetra-quark states.

There are abundant mixing analyses in the literature involving mesons, glueballs and hybrids
utilizing several different methods including
perturbation
theory, relativistic Feynman-Schwinger path integrals for Green's
functions \cite{Simonov}, instantons
\cite{Kochelev}, lattice QCD \cite{Lee} and  effective chiral
approaches \cite{Giacosa}. 
There have also been several meson-glueball phenomenological mixing studies \cite{Burakovsky,Celenza,Close,Li}
in which the meson-glueball interaction is modeled, or simply parameterized,
and then
diagonalized to obtain relations between 
primitive and physical masses.

Central to this work is the mixing between meson and tetra-quark states,
which has not received much
attention.  The few published studies include 
two  diquark, anti-diquark cluster applications with tetra-quark mixings between
either hybrid \cite{Noya} or quarkonium  \cite{Giacosa2} states,
a  meson-meson coupled-channels scattering calculation \cite{er} and a quark model mixing study \cite{Brac} which obtained
an improved  scalar spectrum  
by adjusting phenomenological
parameters.

As detailed in this paper,  meson-tetra-quark mixing is fundamentally
due  to $q \bar{q}$ pair annihilation/formation and entails
the strong confining interaction.  We include this mechanism using our Coulomb gauge (CG) model, which has
been successfully applied to  meson, glueball,
hybrid and tetra-quark states
\cite{Cotanch,Ignacio,Ignacio2,LC2,LCSS,LBC,LC1,LChybrid}. The model
Hamiltonian is obtained from the QCD Coulomb gauge Hamiltonian 
using a few simplifications (see below). In this way, the
original non-perturbative confining interaction can be rearranged
into a calculable effective potential between  color densities.
 We utilize powerful many-body techniques and relativistic field theory in which
 the non-perturbative vacuum is described as a
coherent BCS ground state with quark and  gluon Cooper pairs (condensates). The 
resulting  model retains the key QCD elements and is thus capable of
robust predictions as comprehensively documented in numerous publications 
\cite{Cotanch,Ignacio,Ignacio2,LC2,LCSS,LBC,LC1,LChybrid}.

We only focus  on the mixing between $q \bar{q}$ and $q \bar{q} q \bar{q}$
configurations and, for two reasons,  omit $q \bar{q}g$  hybrid and $gg$  glueball states.
First, we are interested in low mass spectra where 
energy level mixing arguments  imply that
effects from glueballs and hybrid mesons should
not be large since these exotic hadrons have somewhat heavier masses.  Indeed, in our previous
model applications (and others including lattice results) the lightest
hybrid and glueball masses are predicted to be slightly above \cite{Cotanch,Ignacio} and below
\cite{LBC}   2 GeV, respectively.  This is in contrast to
tetra-quark masses which, due to  four different
color configurations, can be much lighter.
In particular, we calculated  \cite{Cotanch,Ignacio2} the lightest
tetra-quark  to be in the color singlet-singlet state with mass closer to 1 rather than 2 GeV. 
The other reason for omitting  quark-hybrid and quark-glueball mixing matrix elements is that for our model Hamiltonian
(see Section 4) the former are perturbative, and thus expected weak, while the latter
entirely vanish (mixing must proceed via higher order intermediate states).  This would also suggest that glueball widths might  not be large, as typically expected,  perhaps even  narrow, consistent with a recent theoretical prediction \cite{bclr}.  The issue of mixing involving gluonic states, however, merits a further  study
which we plan to address in a separate communication.

This paper is organized into seven sections.  In the next section  we detail the
QCD Coulomb gauge Hamiltonian and then, in Section 3, present a perturbative
analysis of the exact Coulomb kernel.  This motivates  our Coulomb gauge model Hamiltonian 
described in Section 4. Meson and tetra-quark mixing is
treated in Section 5 with numerical results 
given in Section 6. Finally, 
key findings and conclusions are summarized in Section 7.

\section{QCD Coulomb gauge Hamiltonian}

The exact QCD Hamiltonian in the Coulomb gauge \cite{T-D-Lee} is
(summation over repeated indices is used throughout)
\begin{eqnarray}
H_{\rm QCD} &=& H_q + H_g +H_{qg} + H_{C}    \\
H_q &=& \int d{\bs x} \Psi^\dagger ({\bs x}) [ -i
{\mbox{\boldmath$\alpha$\unboldmath}} \cdot
{\mbox{\boldmath$\nabla$\unboldmath}}
+  \beta m] \Psi ({\bs x})   \\
H_g &=& \frac{1}{2} \int\!\! d {\bs x} \!\! \left[ \FP^{-1}{\bf
\Pi}^a({\bs x})\cdot \!\!  \FP {\bf
\Pi}^a({\bs x}) +{\bf B}^a({\bs x})\cdot{\bf B}^a({\bs x}) \right] \; \;  \; \; \\
H_{qg} &=&  g \int d {\bs x} \; {\bf J}^a ({\bs x})
\cdot {\bf A}^a({\bs x}) \label{eq:J.A}\\
H_C &=& -\frac{g^2}{2} \int d{\bs x} d{\bs y}\FP^{-1} \rho^a ({\bs x})
 K^{ab}( {\bs x},{\bs y}  ) \FP \rho^b ({\bs y})   \ ,
\end{eqnarray}
where $g$ is the QCD coupling constant, $\Psi$ is the quark field with current
quark mass $m$, ${A}^a =({\bf A}^a, A^a_0)$ are the gluon fields satisfying the
transverse gauge condition, $\bs{\nabla}\cdot{\bf A}^a = 0$ $(a = 1, 2, ... 8)$,
 ${\bf \Pi}^a = -{ \bf E}^a_{tr} $ are the conjugate momenta and 
 \begin{eqnarray}
\mathbf{E}_{tr}^a&=&-\dot{\mathbf{A}}^a + g( 1- {\mbox{$\nabla$\unboldmath}}^{-2}
{\mbox{\boldmath$\nabla$\unboldmath}}{\mbox{\boldmath$\nabla$\unboldmath}} \cdot  )f^{abc} A^{b}_0\mathbf{A}^c  \\
\mathbf{E}^a&=&-\dot{\mathbf{A}}^a-{\mbox{\boldmath$\nabla$\unboldmath}}
A^{a}_0+gf^{abc} A^{b}_0\mathbf{A}^c 
\\
{\bf B}^a &=& \nabla \times {\bf A}^a + \frac{1}{2} g f^{abc} {\bf
A}^b \times {\bf A}^c \ ,
\end{eqnarray}
are the non-abelian chromodynamic fields.
The color densities, $\rho^a({\bs x})$, and quark currents,
${\bf J}^a$, are 
\begin{eqnarray}
\rho^a({\bs x}) &=& \Psi^\dagger({\bs x}) T^a\Psi({\bs x})
+f^{abc}{\bf
A}^b({\bs x})\cdot{\bf \Pi}^c({\bs x}) \\
{\bf J}^a &=& \Psi^\dagger ({\bs x})
\mbox{\boldmath$\alpha$\unboldmath}T^a \Psi ({\bs x}),
\end{eqnarray}
with standard    $SU(3)$
color matrices,  $T^a = \frac{\lambda^a}{2}$, and structure constants, $f^{abc}$. The
Faddeev-Popov determinant, $\FP = {\rm det}(\mathcal M)$, of the
matrix ${\mathcal M} = {\mbox{\boldmath$\nabla$\unboldmath}} \cdot
{\bf D}$ with covariant derivative ${\bf D}^{ab} =
\delta^{ab}{\mbox{\boldmath$\nabla$\unboldmath}}  - g f^{abc} {\bf
A}^c$, is a measure of the gauge manifold curvature and the kernel
in Eq. (5) is given by $K^{ab}({\bs x}, {\bs y}) = \langle{\bs x},
a|{\mathcal M}^{-1} \nabla^2 {\mathcal M}^{-1}  |{\bs y}, b\rangle$.
The Coulomb gauge Hamiltonian is renormalizable, permits resolution
of the Gribov problem, preserves rotational invariance, avoids
spurious retardation corrections, aids identification of dominant,
low energy  potentials and  introduces  only physical degrees of
freedom (no ghosts) \cite{dz}.

The bare parton fields have the  normal mode expansions
(bare quark spinors $u, v$, helicity, $\lambda = \pm 1$, and color
vectors $\hat{{\epsilon}}_{{\cal C }= 1,2,3}$)
\begin{eqnarray}
\label{colorfields1}
 \Psi(\boldsymbol{x}) &=&\int \!\! \frac{d
    \boldsymbol{k}}{(2\pi)^3} \Psi_{{\cal C}} (\boldsymbol{k})  e^{i \boldsymbol{k} \cdot \boldsymbol{x}} \hat{{\epsilon}}_{\cal C}  \\
\Psi_{ {\cal C}} (\boldsymbol{k})   & = & {u}_{\lambda}
(\boldsymbol{k}) b_{\lambda {\cal C}}(\boldsymbol{k)}  +
{v}_{\lambda} (-\boldsymbol{k})
    d^\dag_{\lambda {\cal C}}(\boldsymbol{-k)}   \\
{\bf A}^a({\bs{x}}) &=&  \int \frac{d{\bs{k}}}{(2\pi)^3}
\frac{1}{\sqrt{2k}}[{\bf a}^a({\bs{k}}) + {\bf
a}^{a\dag}(-{\bs{k}})] e^{i{\bs{k}}\cdot {\bs {x}}}  \ \ \
\\
{\bf \Pi}^a({\bs{x}}) &=& \hspace{-.15cm}-i \int
\!\!\frac{d{\bs{k}}}{(2\pi)^3} \sqrt{\frac{k}{2}} [{\bf
a}^a({\bs{k}})-{\bf a}^{a\dag}(-{\bs{k}})]e^{i{\bs{k}}\cdot
{\bs{x}}}  \!,
\end{eqnarray}
with the Coulomb gauge transverse condition, ${\bs k}\cdot {\bf a}^a
({\bs k}) = \\ (-1)^\mu k_{\mu} a_{-\mu} ^a ({\bs k}) =0$. Here
$b_{\lambda {\cal C}}(\boldsymbol{k)}$, $d_{\lambda {\cal
C}}(\boldsymbol{-k)} $ and  $a_{\mu}^a({\bs{k}})$ ($\mu = 0, \pm 1$)
are the bare quark, anti-quark and gluon Fock operators, the latter
satisfying the  transverse commutation relations
\begin{equation}
[a^a_{\mu}({\bs k}),a^{b \dagger}_{\mu'}({\bs k}')]= (2\pi)^3
\delta_{ab} \delta^3({\bs k}-{\bs k}')D_{{\mu} {\mu'}}({\bs k})  \ ,
\end{equation}
with
\begin{equation}
D_{{\mu} {\mu'}}({\bs k}) = \delta_{{\mu}{\mu'}}-
(-1)^{\mu}\frac{k_{\mu} k_{-\mu'}}{k^2}  \  .
\end{equation}

\section{Perturbative expansion}

Before addressing meson and tetra-quark mixing, we first
report a perturbative study of the kernel $K^{ab}(\boldsymbol{x},\boldsymbol{y})$.
Expanding in powers of $g$ yields
\begin{eqnarray}\nonumber
\mathcal{M}^{-1}\nabla^2\mathcal{M}^{-1} & = &
\nabla^{-2} + 2g\nabla^{-2}\mathcal{A}\nabla^{-2}
\\ &&
+3g^2\nabla^{-2}\mathcal{A}\nabla^{-2}\mathcal{A}\nabla^{-2}  + ...   \  ,
\end{eqnarray}
where $\mathcal{A}^{ab}=f^{abc} {\bf A}^c \cdot \boldsymbol{\nabla}$.
Note the first term represents the simple, long-ranged
Coulomb interaction. We have perturbatively calculated the expectation value of the Hamiltonian component with this kernel for $0^{++}$
$q \bar{q}$ states (the Faddeev-Popov terms are also included)
\begin{eqnarray}\nonumber
E_{C} & \equiv & {\langle\Psi^{JPC}|H_C|\Psi^{JPC}\rangle} 
= g^2 E_2^C + g^4 E_4^C + g^6 E_6^C +  ...  \ .
\end{eqnarray}
 In this paper all wavefunction kets, $|\psi>$, have unit norm.
The  leading diagrams corresponding to
$g^2$, $g^4$ and $g^6$ are shown in Fig. 1. The  $g^2$ diagram contributes
\begin{eqnarray}
E_2^C & =& \int d{\boldsymbol q}d{\boldsymbol q}'
\frac{{\cal{F({\boldsymbol q},{\boldsymbol q}')}}}  {{\boldsymbol p}^2} \\
 {\cal{F}} &=& {\cal U}_{\lambda_1}^\dagger
({\boldsymbol q}) {\cal U}_{\lambda_1^{'}}({\boldsymbol
q}'){\cal V}_{\lambda_2^{'}}^\dagger ({\boldsymbol q}') {\cal V}_{\lambda_2}
({\boldsymbol q}) \Phi_{\lambda_1^{'}
\lambda_2^{'}}^{JPC\dag}({\boldsymbol q}')
\Phi_{\lambda_1\lambda_2}^{JPC}({\boldsymbol q}) \ ,  \nonumber
\end{eqnarray}
with ${\boldsymbol p} = {\boldsymbol q} - {\boldsymbol q}'$ and ${\boldsymbol q}$, ${\boldsymbol q}'$ the initial,
 final quark momenta. The dressed spinors, ${\cal U}_{\lambda}, {\cal V}_{\lambda}$, are BCS rotations of the bare spinors  and
wave function details are given in the following sections. The above integration is convergent
and the numerical value for this Coulomb type interaction energy is $E_2^C = 25.6$ MeV.
\begin{center}
\begin{figure}[tbp]
\hspace{.5cm}
\includegraphics[scale=0.67]{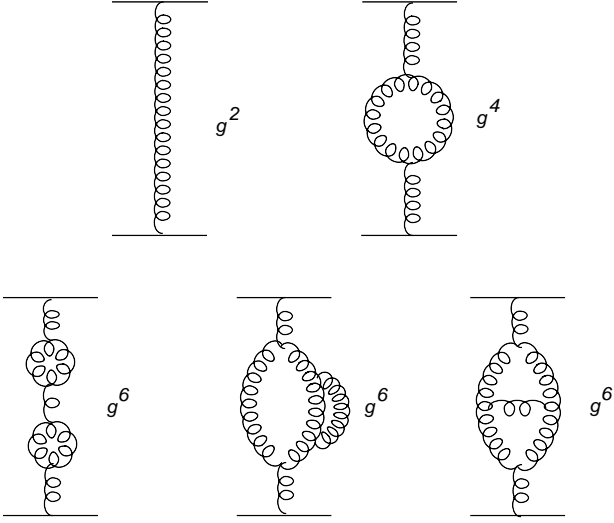}
\vspace{.3cm}
\caption{Diagrams for the kernel expansion to order  $g^6$.}
\end{figure}
\end{center}

The $g^4$ diagram  reduces to 
\begin{eqnarray}
 E_4^C = \int d{\boldsymbol q} d{\boldsymbol q}'
d{\boldsymbol q}_1 \frac{4(1-x_{1}^2){\cal{F({\boldsymbol q},{\boldsymbol q}')}}}{{\boldsymbol p}^2({\boldsymbol
p}-{\boldsymbol q_1})^2\omega({\boldsymbol q}_1)} \ , \;
\end{eqnarray}
where $ x_1 = {\boldsymbol {\hat p}}\cdot {\boldsymbol {\hat q}_1}$.
The momentum integration ${\boldsymbol q}_1$ over the loop 
diverges but replacing the gluon kinetic energy 
$\omega({\boldsymbol q}_1)$ with $\omega({\boldsymbol
q}_1)^{1+\epsilon}$ yields a finite result, $A + B/\epsilon$, for  positive $\epsilon$
which isolates the divergence. 
As shown in Fig. 2, the integrated value scales as
$1/\epsilon$ 
for small  $\epsilon$ which is consistent
with  dimensional renormalization.  Extrapolating the intercept from the linear graph yields the
infinite subtracted renormalized result $E_4^C = A = 13.2 $ MeV. In this minimal subtraction
scheme the coupling $g$ is renormalized to its physical value by absorbing the infinity.
\begin{center}
\begin{figure}[hp]
\hspace{.2cm}
\includegraphics[scale=0.7]{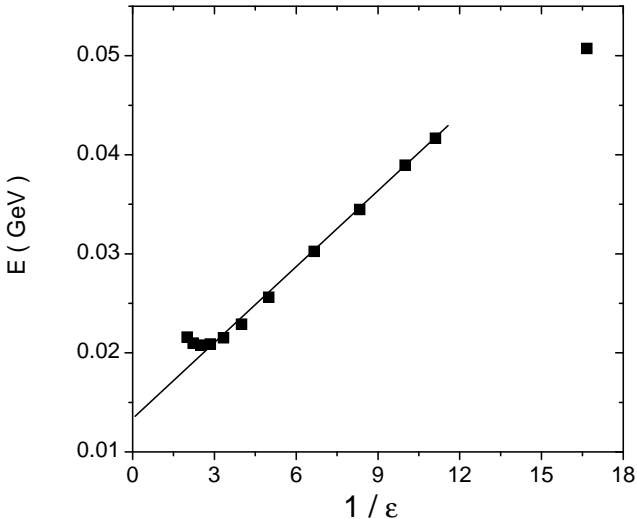}
\vspace{.25cm}
\caption{The interaction energy to order $g^4$ versus $1/\epsilon$.}
\end{figure}
\end{center}

\vspace{-.5cm}

The bottom three diagrams of Fig.~1(from left to right) are proportional to $g^6$, with respective expectation values
\begin{eqnarray}
\!\int \! \! \!  d{\boldsymbol q} d{\boldsymbol
q}' d{\boldsymbol q}_1 d{\boldsymbol
q}_2\frac{16(1-x^2_1)(1-x^2_2){\cal{F({\boldsymbol q},{\boldsymbol q}')}} }
{{\boldsymbol
p}^2({\boldsymbol p}-{\boldsymbol q}_1)^2({\boldsymbol p}-{\boldsymbol q}_2)^2\omega({\boldsymbol q}_1)
\omega({\boldsymbol q}_2)}  \nonumber ,
\end{eqnarray}
\begin{eqnarray}
\int \!\! d{\boldsymbol q} d{\boldsymbol
q}' d{\boldsymbol q}_1 d{\boldsymbol
q}_2 \frac{4(1-x^2_1){\boldsymbol{k}^2_1 (1 - z^2){\cal{F({\boldsymbol q},{\boldsymbol q}')}}}}{{\boldsymbol
p}^2({\boldsymbol p}-{\boldsymbol q}_1)^4({\boldsymbol p}-{\boldsymbol q}_1-{\boldsymbol q}_2)^2\omega({\boldsymbol q}_1)
\omega({\boldsymbol q}_2)} \nonumber  ,
\end{eqnarray}
\begin{eqnarray} \nonumber
\int \!\! \frac{d{\boldsymbol q} d{\boldsymbol
q}' d{\boldsymbol q}_1 d{\boldsymbol
q}_2{\cal{F({\boldsymbol q},{\boldsymbol q}')}}}{{\boldsymbol p}^4
({\boldsymbol p}-{\boldsymbol q}_1)^2({\boldsymbol p}-{\boldsymbol q}_2)^2({\boldsymbol p}-{\boldsymbol q}_1-{\boldsymbol q}_2)^2\omega({\boldsymbol q}_1)
\omega({\boldsymbol q}_2)} ~~~~~~~~ \\
\nonumber \left[\boldsymbol{k}_1\cdot(\boldsymbol{k}_1 - 2\boldsymbol{q}_2) - {\boldsymbol k}_1\cdot {\boldsymbol q}_1(\boldsymbol{k}_1 -2{\boldsymbol q}_2)\cdot
{\boldsymbol q}_1/{\boldsymbol q}_1^2\right]~~~~~~~~ \\
\nonumber \left[\boldsymbol{k}_2\cdot(\boldsymbol{k}_2- 2{\boldsymbol q}_1)-\boldsymbol{k}_2\cdot {\boldsymbol q}_2(\boldsymbol{k}_2- 2{\boldsymbol q}_1)\cdot
{\boldsymbol q}_2/{\boldsymbol q}_2^2\right] ,~~~~~~~~
\end{eqnarray}
where $z = \bs{\hat{k}}_1 \cdot \bs{\hat{q}}_2$, $ x_j = {\boldsymbol {\hat p}}\cdot {\boldsymbol {\hat q}_j}$ 
 and $\bs{k}_j = 2 \bs{p} - \bs{q}_j$ for $j = 1, 2$.
These  loop variable integrations
are also divergent. Using the same renormalization procedure  yields the
respective values $7.15g^6$,
$7.86g^6$ and $0.48g^6$ MeV. Therefore, the sixth order interaction energy is
$E_6^C = 15.5 $ MeV and the series takes the form $E_C =
(25.6g^2 + 13.2g^4 + 15.5 g^6)$ MeV. Since the coefficients are comparable, $g^2$ must be less than  1,
i.e. $\alpha_s = g^2/4 \pi$  $ <$ 0.1, for perturbation theory to be  valid but the strong interaction has
 $\alpha_s$ much larger, so  the  perturbative expansion fails as anticipated.
We therefore seek a calculable confining kernel interaction and, guided by lattice results,
adopt a linear potential specified in the next section.  We note in passing that a subset or class of diagrams
may still be amendable to a perturbative treatment.  Specifically the chain of bubble diagrams (i.e. the $g^4$ and first
$g^6$ diagrams in Fig. 1) seems to be converging.  Also, ladder type diagrams (third $g^6$ diagram) seem much
smaller, in contrast to gluon dressed or self-energy type diagrams (second $g^6$ diagram), and they too may
be convergent.  Hence further perturbative studies should be conducted to determine which parts, if any, of the exact
kernel can be treated as radiative corrections and which parts are responsible for confinement and must
be included non-perturbatively.  This would provide further insight into the nature of confinement and also
for improved QCD approximated interactions. 

\section{Coulomb gauge model Hamiltonian}

Our model's starting point is the Coulomb gauge QCD Hamiltonian,
Eq. (1). In this gauge, the color form of Gauss's law, which
is essential for confinement, is satisfied exactly and can be used to
eliminate the unphysical
longitudinal  gluon fields.
We then make two approximations: 1)
replace the exact Coulomb kernel with a calculable confining potential; 2)
use the lowest order, unit value for the the Faddeev-Popov determinant.  This
defines the CG model
Hamiltonian 
\begin{eqnarray}
H_{\rm CG} &=& H_q + H_g^{\rm CG} + H_{qg} + H_C^{\rm CG}   \\
H_g^{\rm CG} &=& \frac{1}{2} \int d {\bs x}\left[ {\bf
\Pi}^a({\bs x})\cdot {\bf \Pi}^a({\bs x}) +{\bf B}^a({\bs
x})\cdot{\bf B}^a({\bs x})
\right] \label{eq:non-abelian} \\
H_C^{\rm CG} &=& -\frac{1}{2} \int d{\bs x} d{\bs y} \rho^a ({\bs
x}) \hat{V}(\ar {\bs x}-{\bs y} \ar ) \rho^a ({\bs y})   \ .
 \label{model}
\end{eqnarray}
Confinement is described by a  kernel that is a
Cornell type potential, ${\hat V}(r)=-{\alpha_s}/{r}+\sigma r$, where
the string tension, $\sigma=0.135$ GeV$^{2}$, and
$\alpha_s =0.4$ have been previously determined
and set the scale for the calculation.

Next, hadron states are expressed as dressed quark (anti-quark) Fock
operators, $B^{\dag}_{\lambda_1{\cal C}_1}$ ($
    D^{\dag}_{\lambda_2{\cal C}_2}$), acting on the Bardeen-Cooper-Schrieffer (BCS) 
    model vacuum, $|\Omega \rangle$ (see Refs. \cite{Ignacio,LC2} for full details). For the tetra-quark system,
the quark (anti-quark) $cm$ momenta
are   ${\bf q_1}$, ${\bf q_3}$ (${\bf q_2}$, ${\bf q_4}$) and the following
wave function ansatz is adopted
  \begin{eqnarray} 
    |\Psi^{JPC}\rangle = \int \!\!
    \frac{d\boldsymbol{q}_1}{(2\pi)^3} \frac{d\boldsymbol{q}_2}{(2\pi)^3}
    \frac{d\boldsymbol{q}_3}{(2\pi)^3} \Phi^{JPC}_{\lambda_1 \lambda_2
    \lambda_3 \lambda_4}(\boldsymbol{q}_1,\boldsymbol{q}_2,\boldsymbol{q}_3) \; \; \; \;
    \\ \nonumber 
    R^{{\cal C}_1{\cal C}_2}_{{\cal C}_3{\cal C}_4}
    B^{\dag}_{\lambda_1{\cal C}_1}(\boldsymbol{q}_1)
    D^{\dag}_{\lambda_2{\cal C}_2}(\boldsymbol{q}_2)
    B^{\dag}_{\lambda_3{\cal C}_3}(\boldsymbol{q}_3)
    D^{\dag}_{\lambda_4{\cal C}_4}(\boldsymbol{q}_4)|\Omega \rangle \ . ~~~
  \end{eqnarray}
 The
expression for the matrix $R^{{\cal C}_1{\cal C}_2}_{{\cal C}_3{\cal
C}_4}$ depends on the specific color  scheme selected
\cite{Cotanch,Ignacio2}. Here, we focus on the color singlet-singlet scheme,
$[(3 \otimes \bar{3})_{1} \otimes (3 \otimes
\bar{3})_{1}]_1$, where the $q \bar{q}$ pairs couple
to  color singlets, since it gives the lowest mass among 
the four  color representations. This yields $R^{{\cal C}_1{\cal
C}_2}_{{\cal C}_3{\cal C}_4}=\delta_{{\cal C}_1{\cal
C}_2}\delta_{{\cal C}_3{\cal C}_4}$.
The spin part of the wave function is, 
$\langle \frac{1}{2}  \frac{1}{2}  \lambda_1 \lambda_2 |s_A \lambda_A\rangle$
 $\langle  \frac{1}{2}  \frac{1}{2} \lambda_3 \lambda_4 |s_B \lambda_B\rangle$ 
$\langle s_A  s_B \lambda_A  \lambda_B |J 
\lambda_A+\lambda_B\rangle$,
a product of Clebsch-Gordan coefficients. Here
$J$ is the total angular momentum, ${\bs s}_A = {\bs s}_1 + {\bs s}_2$, ${\bs s}_B = {\bs s}_3 + {\bs s}_4$,  and for scalar hadrons all orbital angular momenta, $l_X$,
are zero, consistent with the lowest energy state
(see Section 6 for p-wave pseudo-scalar and vector hadrons).
A Gaussian radial wavefunction is used
(see \cite{Ignacio2} for details)
\begin{eqnarray}
f(q_A,q_B,q_I) =
e^{-\frac{q^2_A}{\alpha^2_A} - \frac{q^2_B}{\alpha^2_B} - \frac{q^2_I}{\alpha^2_I}} \ ,
\end{eqnarray}
with variational parameters $\alpha_A = \alpha_B$ and
$\alpha_I$
determined by minimizing the tetra-quark masses
\begin{eqnarray}
M_{J^{PC}} &=& {\langle\Psi^{JPC}|H_{\rm CG}|\Psi^{JPC}\rangle}  \\
&=& M_{self}+M_{qq}+M_{\bar{q}\bar{q}}+M_{q\bar{q}}+M_{annih} \ ,
\nonumber
\end{eqnarray}
which were  previously calculated \cite{Cotanch,Ignacio2}.
The subscripts indicate the source of each
contribution:  the $q$ and $\bar{q}$ self-energy,
the $qq$, $\bar{q}\bar{q}$ and $q\bar{q}$ scattering, and the
$q\bar{q}$ annihilation, respectively.
Finally, the $q \bar{q}$ meson state is
\begin{eqnarray}
    |\Psi^{JPC}\rangle & = & \int \!\! \frac{d\boldsymbol{k}}{(2\pi)^3} \Phi^{JPC}_{\lambda_1 \lambda_2}(\boldsymbol{k})
    B^{\dag}_{\lambda_1}(\boldsymbol{k}) D^{\dag}_{\lambda_2}(\boldsymbol{-k}) |\Omega \rangle .
  \end{eqnarray}

\section{Meson and tetra-quark mixing}

In this section, we discuss the meson and tetra-quark mixing for
the $J^{PC}=0^{\pm +}$ and $1^{--}$ states. As discussed above, 
only mixing
between  flavored ($u,d,s$) $q \bar{q}$ mesons and tetra-quarks is investigated.
Using the notation, $|q \bar{q}>$ and $|q \bar{q} q \bar{q}>$ for $|\Psi^{JPC}>$, the  mixed state is  given by
\begin{equation}
|J^{PC}\rangle =
a|n\bar{n}\rangle+b|s\bar{s}\rangle+c|n\bar{n}n\bar{n}\rangle+d|n\bar{n}s\bar{s}\rangle,
\end{equation}
where $n\bar{n}=\frac1{\sqrt{2}}(u\bar{u}+d\bar{d})$. The state
$| s\bar{s}s\bar{s}\rangle$ is not included since its
mass is much higher than the meson masses. The coefficients $a,b,c$
and $d$ are determined by diagonalizing the Hamiltonian matrix, in which
the meson-tetra-quark off-diagonal mixing element is (only  $H_C^{\rm CG}$  contributes) 
\begin{equation} \label{def}
M = {\langle q\bar{q}|H_C^{\rm CG}|q\bar{q}q\bar{q}\rangle}
,
\end{equation}
where $| q\bar{q} \rangle$ is $| n\bar{n}\rangle$ or $|
s\bar{s}\rangle$, and $| q\bar{q}q\bar{q} \rangle$ is $|
n\bar{n}n\bar{n}\rangle$ or $| n\bar{n}s\bar{s}\rangle$.
There are six off-diagonal matrix elements however two,
$\langle s\bar{s}|H_C^{\rm CG}|n\bar{n}\rangle$ and $\langle s\bar{s}|H_C^{\rm CG}|n\bar{n}n\bar{n}\rangle$,
vanish and one (see below), $\langle n \bar{n}  n\bar{n}|H^{\rm CG}_C|n\bar{n} s \bar{s}\rangle$, is computed very small.
The remaining   mixing matrix
elements  are, $\langle n\bar{n}|H^{\rm CG}_C|n\bar{n}n\bar{n}\rangle$,
$\langle n\bar{n}|H_C^{\rm CG}|n\bar{n}s\bar{s}\rangle$ and $\langle
s\bar{s}|H_C^{\rm CG}|n\bar{n}s\bar{s}\rangle$. For our model Hamiltonian, there are two types  of
mixing diagrams illustrated in Fig.~3. Because
of color factors,  nonzero mixing only exists   for
$q \bar{q}$ annihilation   between  different singlet
$q \bar{q}$ clusters. The expression for the first diagram in Fig.~3 is
\begin{eqnarray}
&& M_1 = \frac12 \int\!\!\! \; d{\boldsymbol
q}_1^{} d{\boldsymbol q}_2^{} d{\boldsymbol q}_3^{}V(k)
{\cal U}_{\lambda_1}^\dag({\boldsymbol q}_1^{})
{\cal U}_{\lambda'_1}(-{\boldsymbol q}_4) \\ \nonumber &&
{\cal U}_{\lambda_3^{}}^\dag({\boldsymbol q}_3^{})
{\cal V}_{\lambda_2^{}}({\boldsymbol q}_2^{})\Phi_{\lambda_1^{}
\lambda_2^{} \lambda_3^{}\lambda_4^{}}^{JPC\dag}({\boldsymbol
q}_1^{},{\boldsymbol q}_2,{\boldsymbol q}_3^{})
\Phi_{\lambda'_1\lambda_4}^{JPC}(-2{\boldsymbol
q}_4),
\end{eqnarray}
with   ${\boldsymbol q}_4= -{\boldsymbol q}_1 - {\boldsymbol k}$,  ${\boldsymbol
k}={\boldsymbol q}_2+{\boldsymbol q}_3$ and dressed, BCS
 spinors
\begin{eqnarray}
{\cal U}_\lambda = \frac1{\sqrt{2}}\left (
\begin{array}{lcr}
\sqrt{1+sin \, \phi(q)} \\
\sqrt{1-sin \, \phi(q)} \; \;  {\boldsymbol \sigma}\cdot  {\boldsymbol {\hat q}} \\
\end{array}
\right )\chi_\lambda
\end{eqnarray}
\begin{eqnarray}
{\cal V}_\lambda = \frac1{\sqrt{2}}\left (
\begin{array}{lcr}
-\sqrt{1-sin \, \phi(q)}  \; \; {\boldsymbol \sigma}\cdot  {\boldsymbol {\hat q}} \\
\sqrt{1+sin \, \phi(q)} \; \;  \\
\end{array}
\right )\chi_\lambda  \ .
\end{eqnarray}
\begin{center}
\begin{figure}[tp]
\hspace{.5cm}
\includegraphics[scale=0.45]{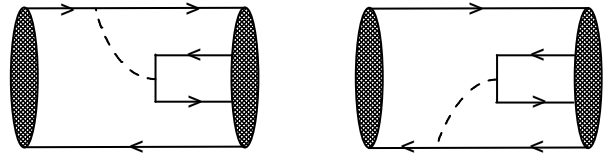}
\caption{Diagrams for the meson, tetra-quark mixing term.}
\end{figure}
\end{center}
  The gap angle, $\phi(q)$, is the solution to the gap equation that
minimizes the energy of the BCS vacuum,
i.e., the vacuum rotated by a Bogoliubov-Valatin transformation
\cite{LC2}.  The
effective confining potential in momentum space is $V(k)$. 
The   second diagram in Fig.~3 yields
\begin{eqnarray}
&& M_2 = \frac12 \int\!\!\! \; d{\boldsymbol
q}_1 d{\boldsymbol q}_2 d{\boldsymbol q}_3 V(k)
{\cal V}_{\lambda_4}^\dag({\boldsymbol q}_4)
{\cal V}_{\lambda'_4}({-\boldsymbol q}_1) \\ \nonumber &&
{\cal U}_{\lambda_3}^\dag({\boldsymbol q}_3)
{\cal V}_{\lambda_2}({\boldsymbol q}_2)\Phi_{\lambda_1
\lambda_2 \lambda_3\lambda_4}^{JPC\dag}({\boldsymbol
q}_1,{\boldsymbol q}_2,{\boldsymbol q}_3)
\Phi_{\lambda_1\lambda'_4}^{JPC}(2{\boldsymbol q}_1) \ .
\end{eqnarray}
\section{Numerical results}

The two Hamiltonian parameters in our model were independently determined
while the wavefunction parameters were obtained variationally.  Because we seek new model masses, the unmixed variational basis states
need not be ones producing a minimal, unmixed mass.  Hence we can use one of the
variational parameters to provide an optimal mixing prediction. We have
selected 
 $\alpha_I$ to exploit this freedom and studied the mixing sensitivity to
 this parameter.

The tetra-quark parity and charge parity  are given by
$P=(-1)^{l_A+l_B+l_I}$ and $C=(-1)^{l_A+s_A+l_B+s_B}$ so
for the lightest, unmixed   $J^{PC} = 0^{\pm +},  1^{--}$ states
\begin{eqnarray}
0^{++} ~~~~~~~~~ l_A=l_B=l_I=0, s_A=s_B=0 ~{\rm or}~ s_A=s_B=1, \nonumber \\
0^{-+} ~~~~~~~~~~~~~~~~~~~~~~~~~~~ l_A=l_B=0, l_I=1, s_A=s_B=1, \nonumber \\
1^{--} ~~~~~~~~~~~~~~~~~~~~~ l_A=l_B=0, l_I=1, s_A=1 ~{\rm or
}~s_B=1. \nonumber
\end{eqnarray}
For the $1^{--}$ p-wave state  we choose $l_I = 1$ 
since this yields a lower mass than states with
 $l_A = 1$ or $l_B = 1$.
Note for the $0^{++}$ state, the spin of the two $q \bar{q}$ clusters 
are both either 0 or 1
and for each the three mixing matrix
elements  versus $\alpha_I$ are shown  in Fig.~4.
The  mixing
term is zero when $\alpha_I$ is zero and then  increases with
increasing $\alpha_I$.
Also, mixing with $s \bar{s}$ states is stronger than with $n \bar{n}$ states.
In particular, for  $\alpha_I=0.2$, the $s_A = s_B = 0$ matrix elements are
$\langle s\bar{s}|H_C^{\rm CG}|n\bar{n}s\bar{s}\rangle =365$ MeV,
$\langle n\bar{n}|H_C^{\rm CG}|n\bar{n}n\bar{n}\rangle =166$ MeV and
$\langle n\bar{n}|H_C^{\rm CG}|n\bar{n}s\bar{s}\rangle =45$ MeV.
\begin{center}
\begin{figure}[t]
\includegraphics[scale=0.75]{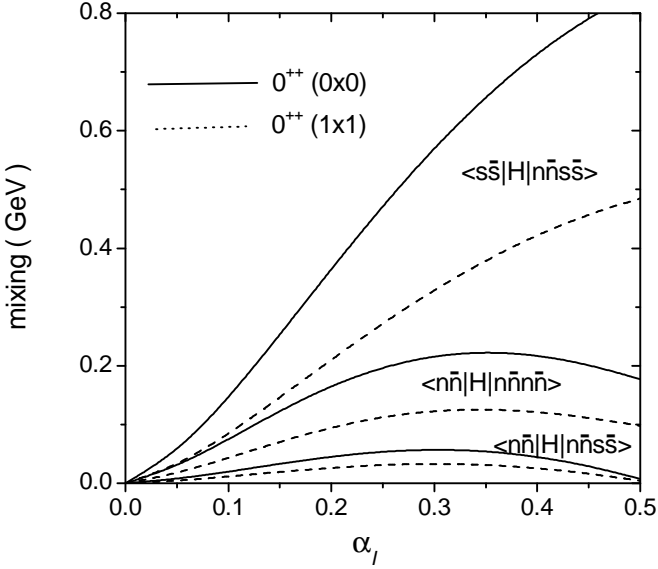}
\caption{The $0^{++}$ mixing matrix elements versus
$\alpha_I$. Solid and dashed lines are for $q \bar{q}$ spin 0 and 1, respectively.}
\end{figure}
\end{center}
\begin{center}
\begin{figure}[bp]
\includegraphics[scale=0.75]{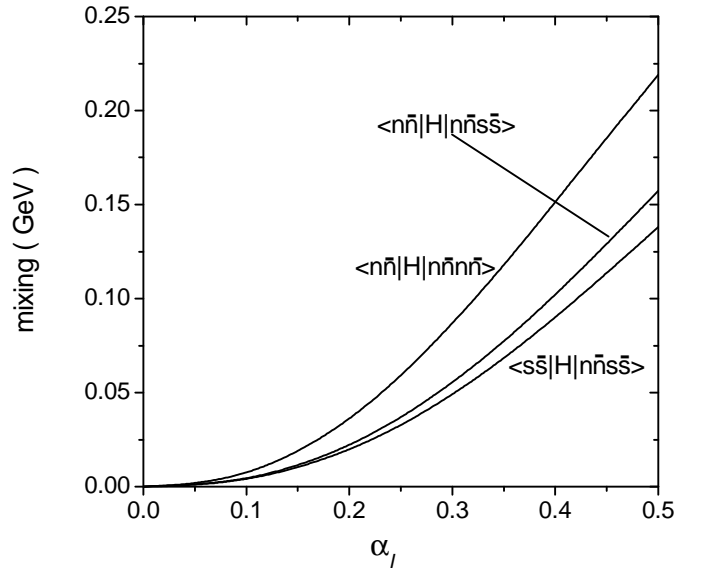}
\caption{The $0^{-+}$  mixing matrix elements  versus
$\alpha_I$.}
\end{figure}
\end{center}

Figure~5 shows the mixing versus $\alpha_I$ for $0^{-+}$ states.
Again, all mixing terms are  zero for  $\alpha_I = 0$ and then
increase with increasing $\alpha_I$. In contrast
to the $0^{++}$ result,  $\langle n\bar{n}|H_{\rm
CG}|n\bar{n}n\bar{n}\rangle$ now has the largest value. The
value $\alpha_I = 0.5$ yields
reasonable  $\eta$ and $\eta{'}$ masses with mixing 
elements  $\langle n\bar{n}|H_C^{\rm
CG}|n\bar{n}n\bar{n}\rangle= 219$ MeV, $\langle n\bar{n}|H_C^{\rm
CG}|n\bar{n}s\bar{s}\rangle= 157$ MeV and $\langle s\bar{s}|H_C^{\rm
CG}|n\bar{n}s\bar{s}\rangle=138$ MeV.

For the $1^{--}$ states a novel mixing result was obtained.
The mixing matrix elements were   again 0 for $\alpha_I = 0$
but, and very interesting, also  essentially 0 for all values of 
$\alpha_I$.  Our model therefore predicts minimal flavor  mixing 
for vector mesons which 
would explain  the
known $\omega/\phi$ ideal mixing.  Related, weak mixing also provides a good spectrum description
since the  pure $n \bar{n}$ and $s \bar{s}$ states were previously   in good agreement \cite{LC2,LCSS} with observation.

As mentioned above, the purely tetra-quark matrix element,
$\langle n\bar{n}n\bar{n}|H_{\rm CG}|n\bar{n}s\bar{s}\rangle$, 
was calculated to be small since only annihilation diagrams contribute.  Its value was  only a few MeV in magnitude for any
$\alpha_I$ and thus has no appreciable effect in this
study.

With the calculated matrix elements and  previously
predicted unmixed meson and tetra-quark masses \cite{Cotanch,Ignacio2,LC2},
 the complete Hamiltonian matrix was diagonalized to obtain the expansion  coefficients and masses
for the corresponding eigenstates.   Using   $\alpha_I=0.2$, the results for $0^{++}$ states are
compared in Table 1 to
 the  observed \cite{PDG} lowest six $0^{++}$ states. Noteworthy, after mixing, the
$\sigma$ meson mass is shifted from 848 MeV to 776 MeV and the strange
scalar meson mass also decreases from 1297 MeV to 1006 MeV, now
close to the experimental value of 980 MeV. Meson and
tetra-quark mixing clearly improves the model predictions as the masses of the other
$f_0$ states are also in better agreement with data.  Figure 6 illustrates the over all
improved description that mixing provides for the $f_0$ spectrum.
New structure insight has also been obtained from the
coefficients, with the  predictions that the $\sigma / f_0(600)$ is predominantly a mixture 
of $n\bar{n}$ and $n\bar{n}n\bar{n}$ states while the $f_0(980)$
 consists mainly of $s\bar{s}$ and
$n\bar{n}s\bar{s}$ states.
\begin{center}
\begin{figure}[bp]
\includegraphics[scale=0.45]{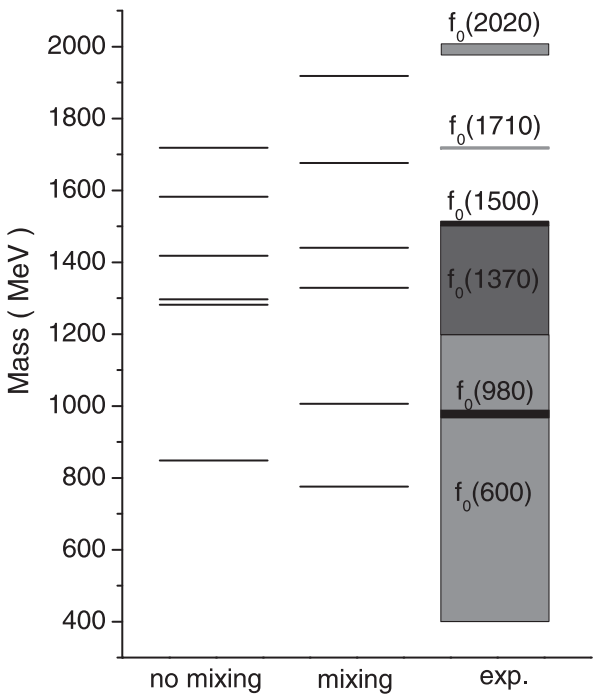}
\caption{Unmixed and mixed $f_0$ spectrum compared to data.}
\end{figure}
\end{center}

\vspace{-1cm}

Table 2 lists the masses and coefficients for the $0^{-+}$ states for $\alpha_I = 0.5$. Again,  mixing lowers (raises) the predicted mass for states relative to unmixed $q \bar{q}$ ($q \bar{q} q \bar{q}$) states.  All mixed
hadron masses are  closer to measurement than the unmixed ones,
except  the most massive state which presumable could further mix with omitted heavier configurations. 
Note from the expansion coefficients that flavor mixing is again weak (i.e. $n \bar{n}$ with $s \bar{s}$) and
less likely than meson, tetra-quark mixing having the same flavors.

\section{Summary and conclusions}

We have applied the established CG model to study $q\bar{q}$ and $q\bar{q}q\bar{q}$
mixing for the low-lying $0^{++}$, $0^{-+}$ and $1^{--}$ spectra. 
In general, mixing effects are significant and provide an improved hadronic description.
As important, our findings clearly document that mixing is necessary
for a complete understanding of  scalar and pseudo-scalar hadrons.

The mixed $0^{++}$ states are a superposition of six 
states with  coefficients  obtained by diagonalizing the
$H_{CG}$ Hamiltonian which decreases the mass for states dominated by $q {\bar q}$ components while increasing
 those predominantly having tetra-quark configurations.
The resulting $f_0$ mass spectrum is in good
agreement with observation.

Mixing is not as large for the $0^{-+}$ spectrum, 
therefore the mass shifts are smaller. Again after mixing, predominantly $q \bar{q} q \bar{q}$ states increase in mass while the  $q {\bar q}$ dominated masses decrease. All  mixed states are closer to 
measurement except  the heaviest which might be further corrected via
mixing with omitted higher configurations.  It is noteworthy that the CG model provides sufficient
flavor mixing to produce reasonable masses for historically challenging $\eta, \eta'$ system. 

Significantly, mixing is calculated to be weak for the $1^{--}$ states.
Therefore,  minimal flavor mixing for vector mesons follows
naturally from our model, consistent with the known $\omega$/$\phi$ ideal mixing.

Finally, we performed a perturbative investigation of the exact QCD Coulomb gauge kernel
to order $g^6$.  As expected, a series expansion in $g$ does not converge, however a subset
class of diagrams might be amendable to a perturbative treatment and further study is recommended.
Future work should also address mixing applications to other $I^G (J^{PC})$ states, especially
glueball and hybrid mesons with explicit gluonic degrees of freedom.  Determining the level
of mixing for these  exotic systems will be important for finally
establishing their existence.

$\\$

\noindent
{\it Acknowledgements.}
The authors are very appreciative for the assistance and advice from 
F.  J. Llanes-Estrada. Supported in part by  U. S. DOE grants
DE-FG02-97ER41048 and DE-FG02-03ER41260.

\onecolumn
\begin{table}[t]
    \caption{Mixing coefficients and masses in MeV for $0^{++}$ states.}
    \begin{center}
    \begin{tabular}{ccccccc}
    \hline \hline \noalign{\smallskip}
  & $|n\bar{n}>$ & $|s\bar{s}>$ & $|n\bar{n}n\bar{n}>_1$ &
$|n\bar{n}n\bar{n}>_2$ & $|n\bar{n}s\bar{s}>_1$& $|n\bar{n}s\bar{s}>_2$ \\
    \hline \noalign{\smallskip}
no mixing & 848 & 1297 & 1282  & 1418 & 1582 & 1718  \\
mixing & 776 & 1006 & 1329 & 1440 & 1676 & 1918 \\
exp. & $f_0(600)$ & $f_0(980)$ & $f_0(1370)$ & $f_0(1500)$ & $
f_0(1710)$ & $f_0(2020)$ \\
& 400 - 1200 & $980 \pm 10$ & 1200 - 1500 & $1507 \pm 5$ & $ 1718\pm
2$
& $1992 \pm 16$ \\
    \hline \noalign{\smallskip}
coeff.  & a & b &  $c_1$ & $c_2$ & $d_1$ & $d_2$  \\
$f_0(600)$  & 0.936 & -0.075 & 0.263  & 0.216 & 0.030 & -0.007 \\
$f_0(980)$ & 0.057 & 0.818  & -0.022 & -0.017& 0.549 & -0.156 \\
$f_0(1370)$ & -0.308& 0.045  & 0.922  & 0.228 & 0.008 & -0.003 \\
$f_0(1500)$ & -0.139& 0.017  & -0.282 & 0.949 & 0.006 & -0.003 \\
$f_0(1710)$ & -0.031& -0.240 & -0.001 & -0.002 &0.582 & 0.776 \\
$f_0(2020)$ & -0.063& -0.514 & -0.002 & -0.006& 0.599 & -0.610 \\
    \hline \hline
    \end{tabular}
    \end{center}
\end{table}

\begin{table}
    \caption{Mixing coefficients and masses in MeV for $0^{-+}$ states.}     
     \begin{center}
    \begin{tabular}{ccccc}
    \hline \hline \noalign{\smallskip}
 & $n\bar{n}$ & $s\bar{s}$ & $|n\bar{n}n\bar{n}>$ &
 $|n\bar{n}s\bar{s}>$ \\
    \hline \noalign{\smallskip}
no mixing & 610 & 1002 & 1252  & 1552  \\
mixing & 531 & 970 & 1316 & 1598 \\
exp. & $\eta$ & $\eta'$ & $\eta(1295)$ & $\eta(1405)$ \\
& $547.51\pm 0.18$ & $957.78 \pm 0.14$ & $1294 \pm 4$ & $1409.8 \pm
2.5$ \\
    \hline \noalign{\smallskip}
coeff.  & a & b &  c & d  \\
$\eta$  & 0.951 & -0.046 & 0.279  & 0.126  \\
$\eta'$ & 0.032 & 0.973  & -0.046 & 0.223 \\
$\eta(1295)$ & -0.289& 0.036  & 0.953  & 0.080 \\
$\eta(1405)$ & -0.108& -0.222  & -0.105 & 0.963 \\
    \hline \hline
    \end{tabular}
    \end{center}
\end{table}

\end{document}